\documentclass[prl,twocolumn,superscriptaddress]{revtex4}
\usepackage{graphicx}
\begin{document}
\title{Observations of spontaneous oscillations in simple two-fluid networks}

\author{Brian D. Storey}
\affiliation{Franklin W. Olin College of Engineering, Needham MA 02492}
\author{Deborah V. Hellen}
\affiliation{Franklin W. Olin College of Engineering, Needham MA 02492}
\author{Nathaniel J. Karst}
\affiliation{Babson College, Babson Park, MA 02457}
\author{John B. Geddes}
\affiliation{Franklin W. Olin College of Engineering, Needham MA 02492}

\date{\today}

\begin{abstract}
We investigate the laminar flow of two-fluid mixtures inside a simple network of inter-connected tubes. The fluid system is comprised of two miscible Newtonian fluids of different viscosity which do not mix and remain as nearly distinct phases. Downstream of a diverging network junction the two fluids do not necessarily split in equal fraction and thus heterogeneity is introduced into network. We find that in the simplest network, a single loop with one inlet and one outlet, under steady inlet conditions the flow rates and distribution of the two fluids within the network loop can undergo persistent spontaneous oscillations. We develop a simple model which highlights the basic mechanism of the instability and we demonstrate that the model can predict the region of parameter space where oscillations exist. The model predictions are in good agreement with experimental observations.
\end{abstract}
\maketitle

\section{Introduction}
In  piping networks where the fluid is comprised of multiple phases or constituents it has been observed that the phase distribution within the network may exhibit unsteady or non-unique flow.  At the micro-scale, the flow of droplets or bubbles through microfluidic networks can demonstrate bistabilty, spontaneous oscillations, and non-linear dynamics ~\cite{Jousse2006,Schindler2008,Fuerstman2007,Prakash2007, Joanicot2005}.  On the macro-scale,  models of magma flow in lava tubes have shown the existence of multiple solutions on the pressure-flow curve which can lead to spontaneous  oscillations in the flow ~\cite{Helfrich1995, Wylie1999}. A well-studied network that can exhibit complex dynamic behavior is microvascular blood flow where August Krogh first noted the heterogeneity of blood flow in the webbed feet of frogs in 1921~\cite{Krogh:1921aa}.
 Simulations, analysis, and experiments with microvascular networks have demonstrated the possibility of spontaneous oscillations in flow rates and hematocrit distribution though direct validation between model and experiment is lacking~\cite{Kiani:1994aa,Carr:2000aa,Geddes:2007,Pop:2007aa,Shevkoplyas, Owen, Pozrikidis}.

There are two fundamental phenomena in two fluid networks which differ from their single fluid counterparts. The first effect is that the effective viscosity in a single pipe (or vessel) depends upon the fraction of the different fluids in the pipe. The second effect is that  the phase fraction after a diverging junction  may be different in the two downstream branches. Such phase separation at a node exists in numerous systems. In microvascular blood flow, Krogh introduced the term ``plasma skimming''  in order to explain the disproportionate distribution of red blood cells at vessel bifurcations ~\cite{Krogh:1921aa,Bugliarello:1964aa,Chien:1985aa,Dellimore:1983aa,Fenton:1985aa,Klitzman:1982aa,Pries:1989aa}. Another example are industrially relevant gas-liquid flows where extensive experimental work  has been conducted ~\cite{Azzopardi1999,Azzopardi1994,lahey1986}.

Recent work by our group has focused on simple networks containing two miscible Newtonian fluids of differing viscosities. This fluid system provides controllable laboratory experiments and the simple network geometries are amenable to analysis.
Through theory and experiment, we have  shown that the existence of phase separation at a single junction
and non-linear mixture viscosity
in this system can lead to multiple stable equilibrium states within the network~\cite{geddes2010a,karst2013}.
 We recently conducted a theoretical study  of dynamics in  networks with this fluid system \cite{karst2014}. We used a combination of analytic and numerical techniques to identify and track saddle-node and Hopf bifurcations through the large parameter space. We found  predictions of sustained spontaneous oscillations in the flow rates internal to the network for steady inlet conditions.

In this paper we build upon our prior work and  experimentally verify predictions on 
the existence of spontaneous oscillations  within
 simple two-fluid networks.  The fluid system is two miscible Newtonian fluids of differing viscosities and densities such that there is stratified flow within each  tube \cite{karst2013}. The network is a only single loop with one inlet and one outlet. The inlet to the network loop is held steady, yet we observe under certain conditions that the contents of the branches inside the loop are unsteady.
We develop a simple model that explains the underlying mechanism of this instability and we 
demonstrate that the model is able to accurately predict the region of parameter space where oscillations exist. 

\section{Experiments}
\begin{figure}
\begin{center}
\includegraphics[width=3.3in]{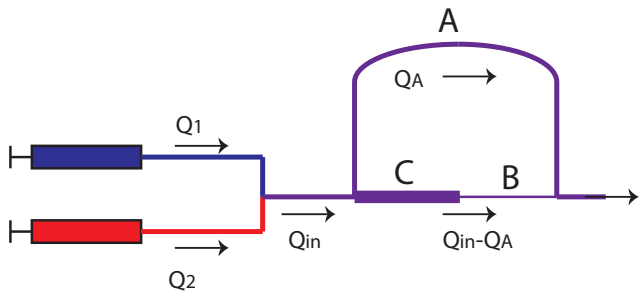}
\end{center}
\caption{Top view schematic of the experimental setup (gravity points into the page). Syringe pumps push two different fluids at controlled rates. Inlet pump 1 contains water while inlet pump 2 contains a viscous glycerol solution. The two fluids merge into a stratified flow before entering a small three vessel network comprised of a single loop.}
\label{fig:schematic}
\end{figure}

A top view schematic of the physical system is shown in Fig.~\ref{fig:schematic} (gravity points into the page).
This network represents perhaps the simplest case where the flow
 internal to the network is not fully determined by the inlet conditions.
Two syringe pumps supply source fluids at a controlled and steady flow rate. Inlet pump 1 contains water (denoted as fluid 1) and inlet pump 2 contains an aqueous glycerol solution (fluid 2). The mass fraction of glycerol in fluid 2 is measured  to set the desired viscosity. Circular tubing (1.6 mm inner diameter) from the two inlet pumps meet at the inlet junction where the density difference of the two fluids is sufficient to create a strongly stratified flow. The dense viscous solution is observed to flow along the lower half of the tube and the water on top. Food dye is added for visualization. The inlet tube then approaches the inlet T-junction to the network comprised of a single loop (see Fig.~\ref{fig:schematic}). The inlet flow rates on both pumps are set to $Q=1$ ml/min which provides an inlet volume fraction of fluid 2 of  $\Phi_{\mathrm{in}}=0.5$.  The T-junction has outlet branches which are 90 degrees relative to each other and the orientation of this diverging  node as shown in Fig.~\ref{fig:schematic} is important for the results we obtain \cite{karst2013}.

The content of vessel C is monitored by imaging every 5 seconds. Imaging allows us to monitor the location of the interface between the two fluids, and thus infer the relative fraction of the two fluids in the tube. If the flow rate inside the loop oscillates then the fraction of the two fluids  in each tube will also change with time. Thus, by simply monitoring the interface location between the two fluids we are able to determine whether the flow is steady or not.

A disadvantage of syringe pumps is that they inherently introduce periodic noise~\cite{lee2012}. For our system the pumps introduce a low amplitude fluctuation with a period of approximately 45 seconds  which appears in all experiments and controls. In order to ensure that any dynamics we observe  are not driven by the pumps we also conduct some experiments with a constant pressure source. We use a pressure regulator and gravity to set a steady inlet pressure from  a reservoir of each inlet fluid. The inlet hydraulic resistances to the network are  tuned to  match the viscosity ratio of the two fluids to order to  set an inlet volume fraction of  $\Phi_{\mathrm{in}}=0.5$.   The disadvantages of the pressure system is that we have only passive control over the flow rate and  as the reservoir drains the applied pressure  decreases very slowly with time. While the resulting dynamics would be expected to be different for constant pressure or constant flow conditions, the underlying existence of instability is identical with either driving.

The sizing of the lengths and diameters in the network were guided by the model predictions  in our previous work \cite{karst2014}. The diameters of the cylindrical vessels was set to $d_A=0.8~\mathrm{mm} = 1/32~\mathrm{in}$, $d_B=0.51~\mathrm{mm} = 1/50~\mathrm{in}$, and $d_C=3.2~\mathrm{mm} = 1/8~\mathrm{in}$. The length of the vessels are varied in each experiment, but typical lengths are on the order of 100 mm.

\section{Model}
A complete model of this system was presented in our previous paper based on  a 1D wave equation for the volume fraction in each vessel  ~\cite{karst2014}.
In the model, the boundary condition to the vessels downstream of the inlet  diverging node  are provided by   a constitutive law which states the phase separation function. For our stratified flow system, this phase separation
behavior  was measured in  prior work ~\cite{karst2013}. The phase separation function provides the volume fraction
in the downstream tubes as a function of the flow into  and out of the node, $Q_{\mathrm{in}}$ and $Q_A$.
When this system of convection equations is linearized, the propagation of fluid through the system manifests itself as a delay and we arrive at a set of state-dependent delay equations.
In network problems if there is a change at the inlet node it takes time for that change to propagate through 
to the exit.

To close this system of convection equations, we must consider the pressure drop, $\Delta P$, across any vessel which is proportional to the flow  $\Delta P_i  = Q_i R_i$. The  hydraulic resistance,  $R_i$, is computed  through Poiseuille's law
\begin{equation}
R_i(t) =  \frac{\bar{\mu}_i(t)}{\mu_1} r_i; ~~~ r_i = \frac{128 \ell_i  {\mu}_1}{  \pi d_i^4}.
\end{equation}
where $r_i$ is the nominal resistance and $\bar{\mu}_i$ is the integrated value of effective viscosity over the tube's length $l_i$.
The local effective viscosity depends upon the volume fraction of the two fluids in the tube.
Summing pressure drops around the  loop provides an equation for the flow inside the network's single loop which is,
\begin{equation}
{Q_A(t) \over Q_{\mathrm{in}} } = {R_B(t) + R_C(t)  \over R_A(t) + R_B(t) + R_C(t)}.\label{eqn:flow}
\end{equation}
Since the flow is incompressible, this flow equation must be satisfied at each instance in time.
In dimensionless terms this convection model depends upon  the ratio of the nominal resistances, $r_C/r_B$ and $r_A/r_B$,   the ratio of the volume of the vessels, $V_A/V_C$ and $V_B/V_C$,  the viscosity contrast $\mu_2/\mu_1$, and  $\Phi_{in}$.

In our previous theoretical paper, we found that the region of parameter space where the convective model shows instability is dominated by the case  when the diameter of vessel C is large relative to all others~\cite{karst2014}. The large diameter introduces a simplifying limit where
$r_C/r_B \rightarrow 0$, $V_A/V_C \rightarrow 0$, and $V_B/V_C \rightarrow 0$. When C has a large diameter that vessel's resistance does not influence the flow equation, Eq. \ref{eqn:flow}. Further, the time delays associated with flow through vessels B and A are so short that they can be assumed instantaneous with respect to changes in C. In this limit, vessel C sets the time delay for the system and vessels A and B are assumed to always be in quasi-equilibrium - namely the contents of those two vessels are uniform along the length. 
These assumptions remove the need for the  full convective flow
equations and instead yields a simple implicit iterative map,
\begin{equation}
{Q_A(t) \over Q_{\mathrm{in}}  }= {R_B(t-\tau_C)  \over R_A(t) + R_B(t-\tau_C)}.
\label{eqn:flow_map}
\end{equation}
The resistance of branch B therefore lags that in A by the time delay which is set by the flow through vessel C,  $\tau_C = V_C/Q_C$. Note that both $R_A$ and $R_B$ depend upon the contents of those vessels respectively and therefore the flow $Q_A$.
While our equation is implicit, it is similar to the classic iteration equations in discrete dynamical systems.
 The current value of $Q_A$ is substituted into this implicit
 algebraic equation to determine the value of $Q_A$  at the  time, $\tau_C$, later.
If the value of $Q_A$ converges upon successive iterations of the map  then system is stable.
The value of the delay  time, $\tau_C = V_C/Q_C$, has no impact on the system stability but would impact the resulting frequency of oscillations if the flow were unstable.  This model will be referred to as the iterative map throughout.

The iterative map assumes that vessel C is critical for setting the time delay between the state of vessels A and B, while vessel B is critical for setting the resistance in the B-C branch. If there is a change at the inlet T-junction it takes time for that change to propagate to vessel B. While the change is propagating through C, the flow inside the loop does not change since the resistance of C is unimportant. As soon as the change enters vessel B, the change propagates quickly (relative to the delay in C) and instantaneously changes the resistance in B, thus feeding back to potentially change the state of the network flow. In the iterative map, only the parameters, $r_A/r_B$, $\mu_2/\mu_1$, $\Phi_{in}$ enter the problem.

The stability of the iterative map model can be determined readily \cite{Strogatz}.
Here we focus on locating any bifurcations in which the equilibrium flow loses stability to a period-2 oscillation.
 Following standard procedures, a little algebra
 reduces this criterion to $d F/ dQ_A<0$, where the stability function $F$ is defined by,
\begin{equation}
F = \frac{\Delta P_A \Delta P_B}{\left(1-\frac{Q_A}{Q_{\mathrm{in}}}\right)^2}.
\end{equation}

\begin{figure*}
\begin{center}
\end{center}
a) \includegraphics[width=2.1in]{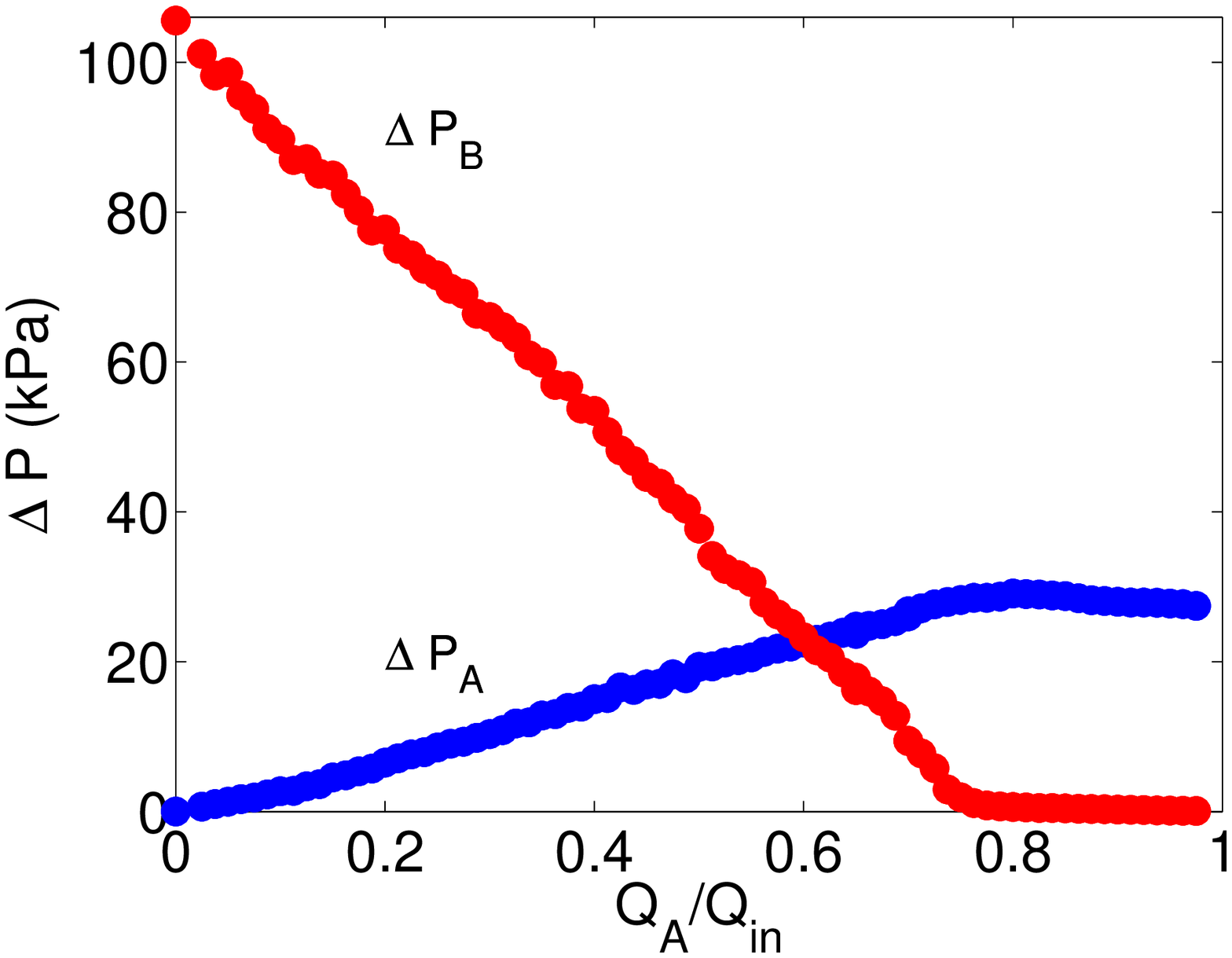}
b) \includegraphics[width=2.1in]{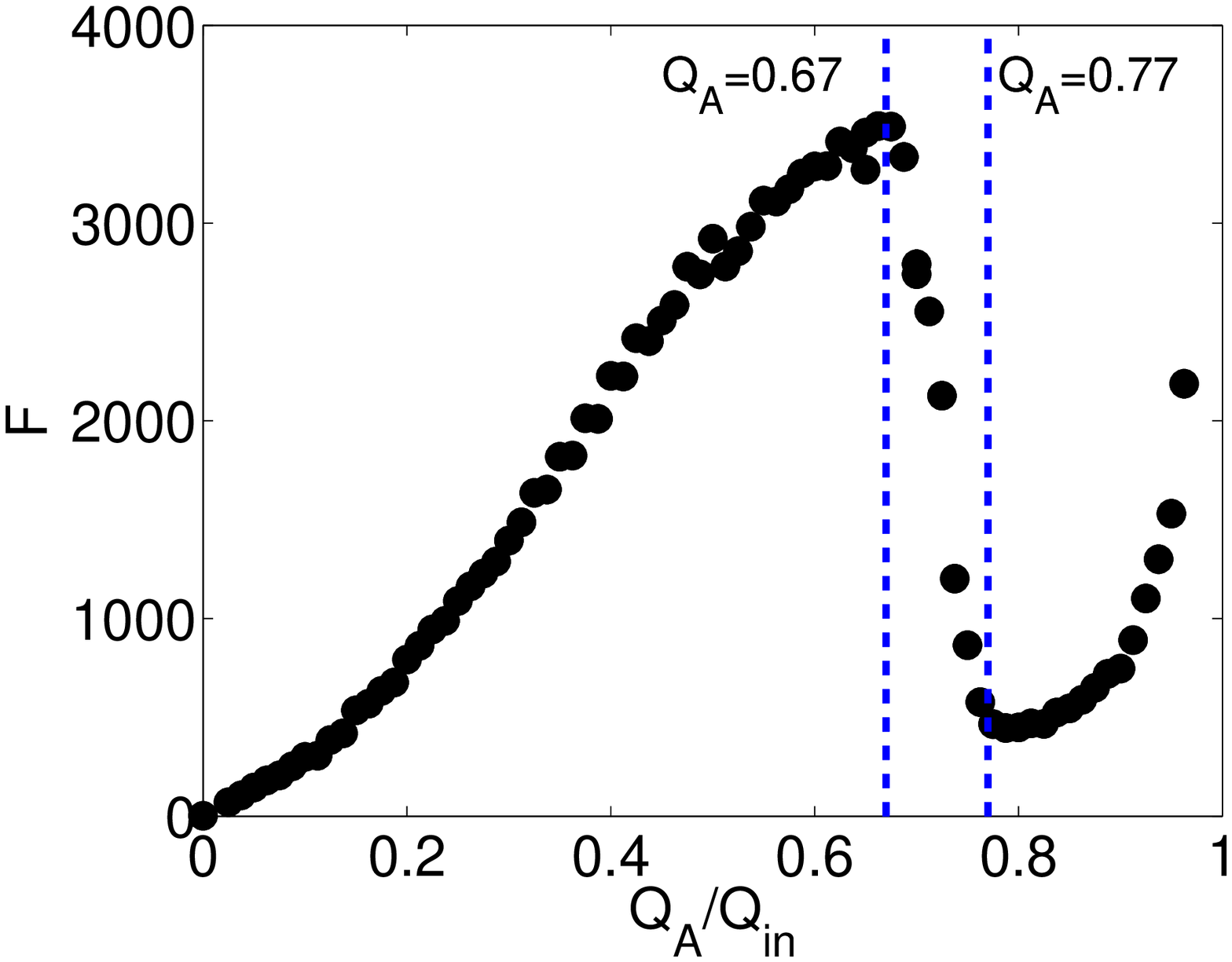}
c) \includegraphics[width=2.1in]{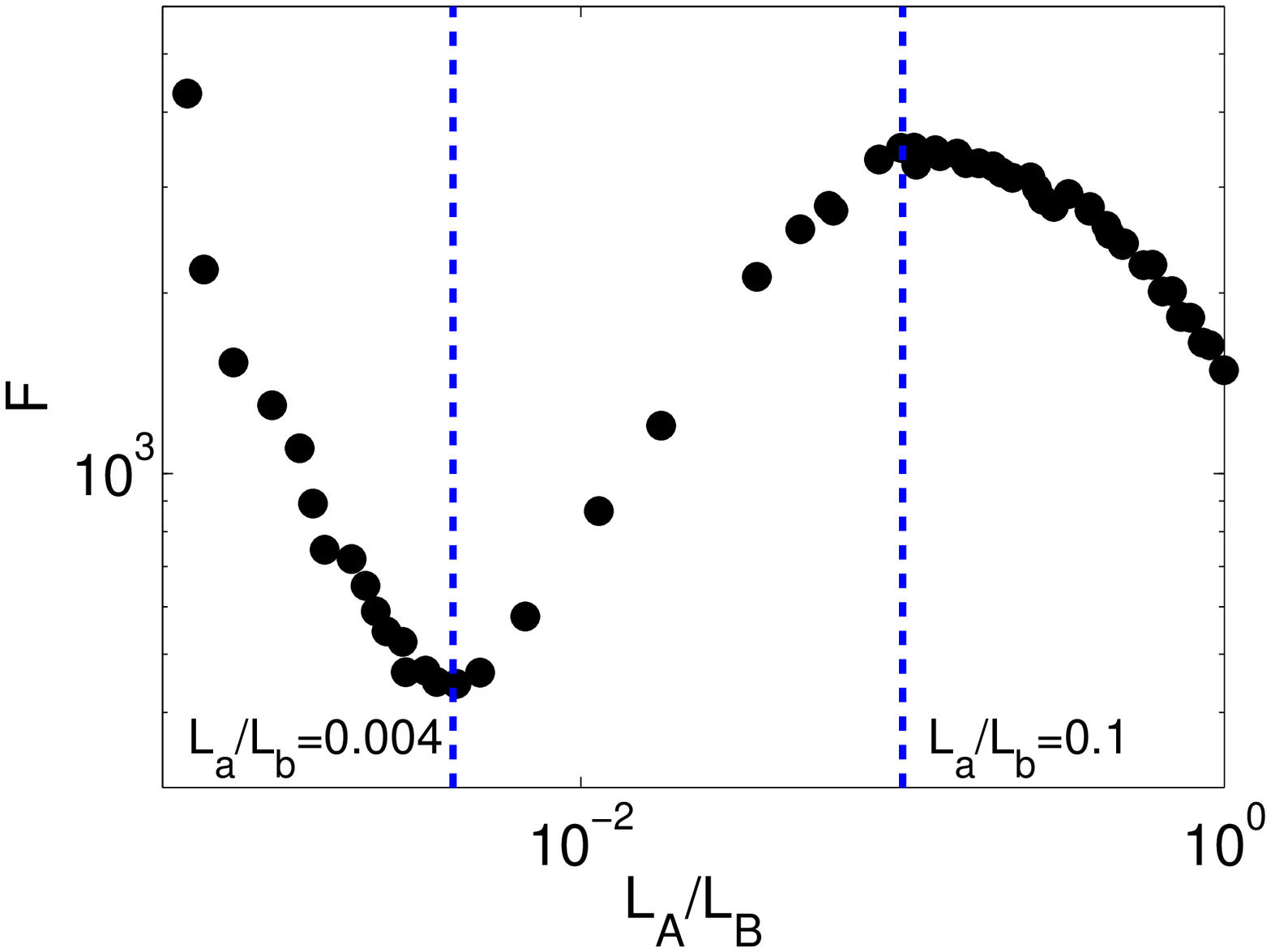}\\
\bigskip
d) \includegraphics[width=1.5in]{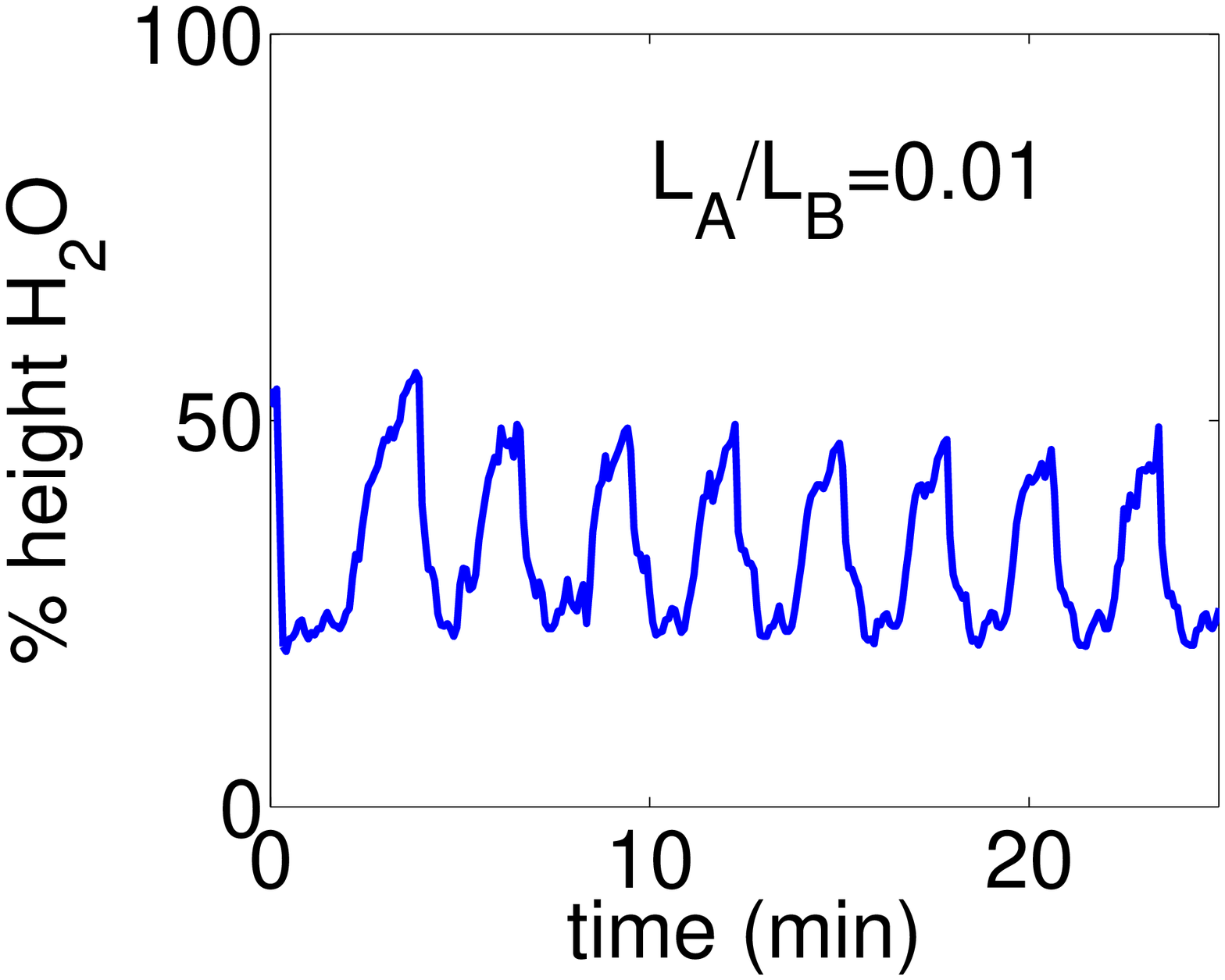}
e) \includegraphics[width=1.5in]{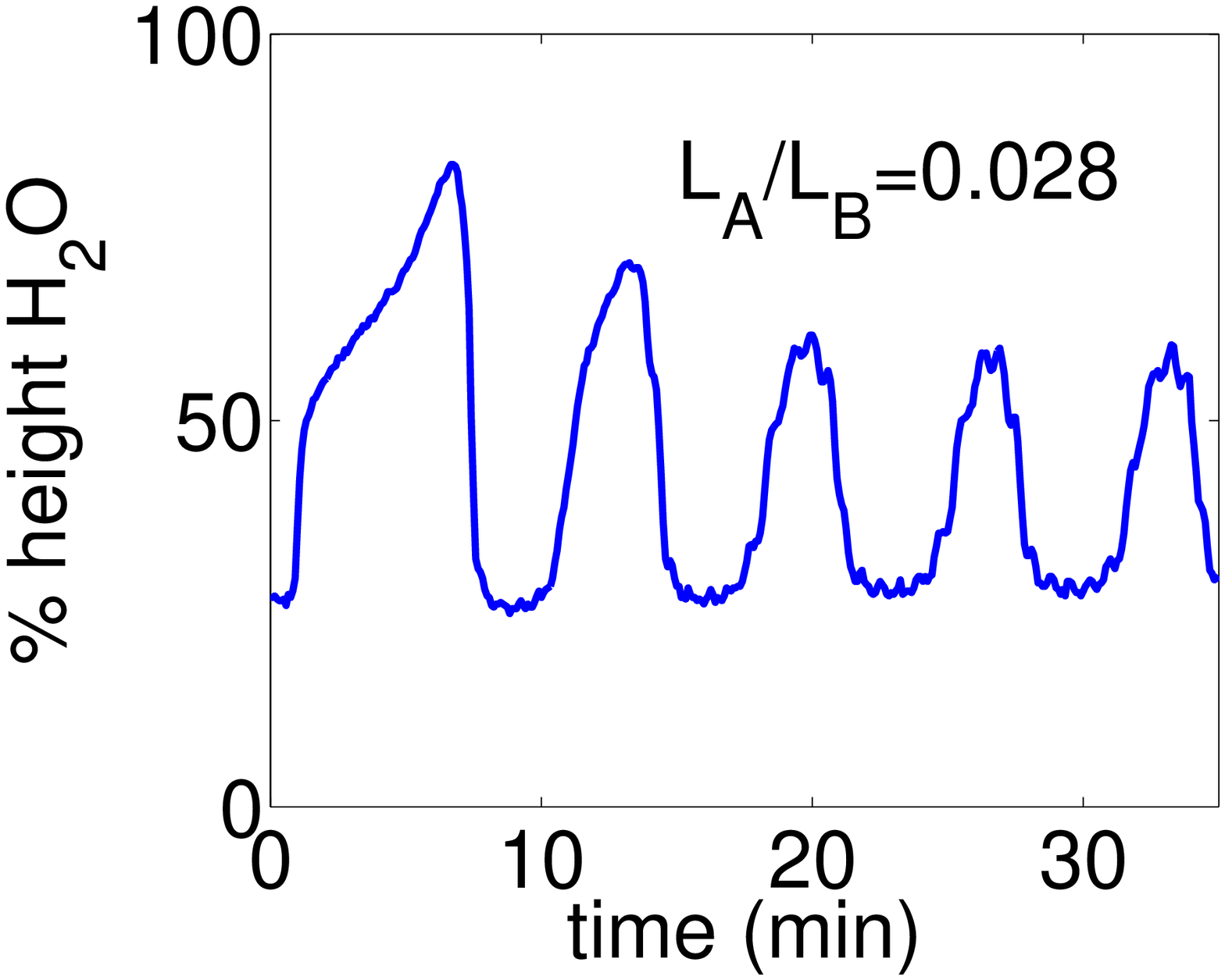}
f) \includegraphics[width=1.5in]{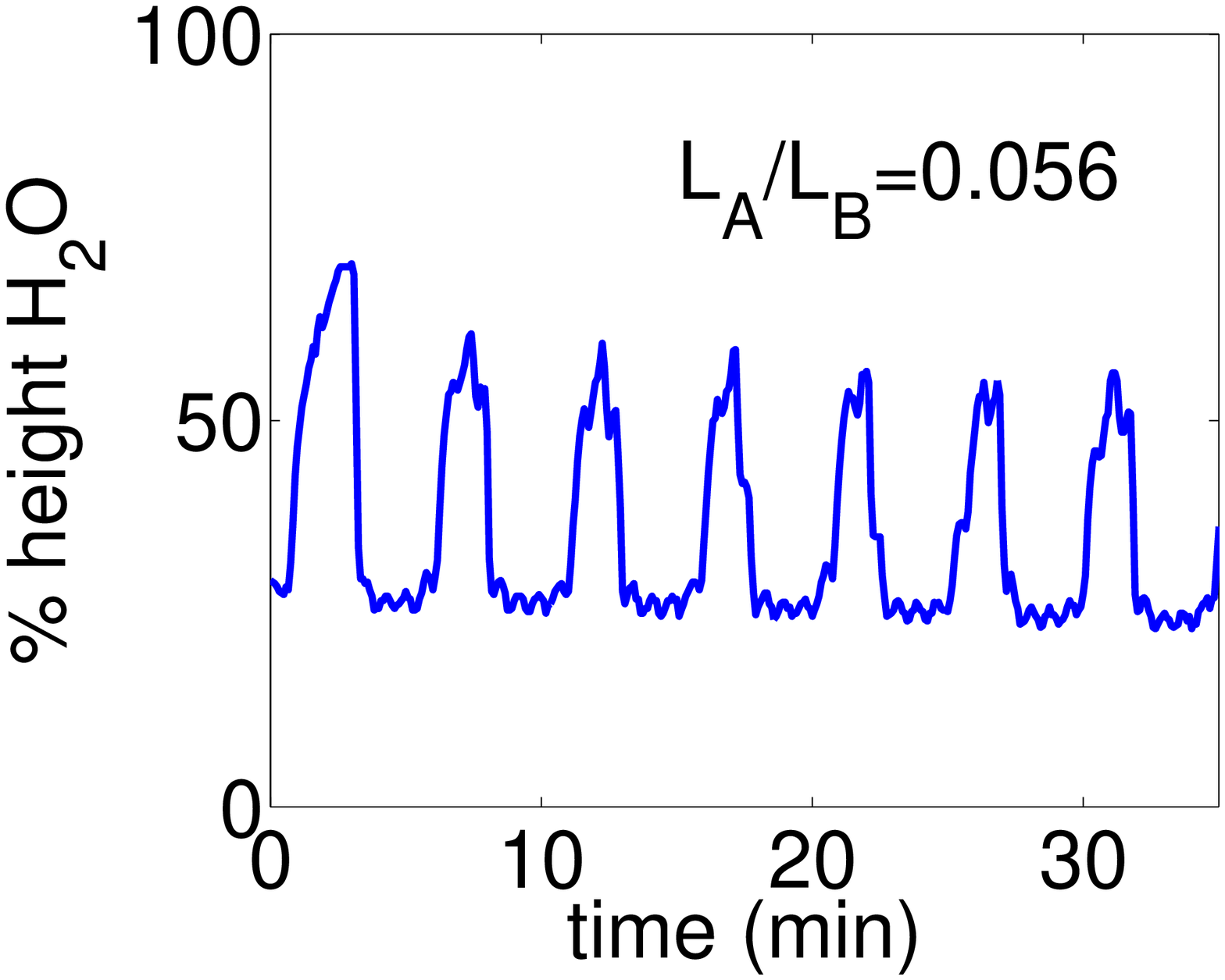}
g) \includegraphics[width=1.5in]{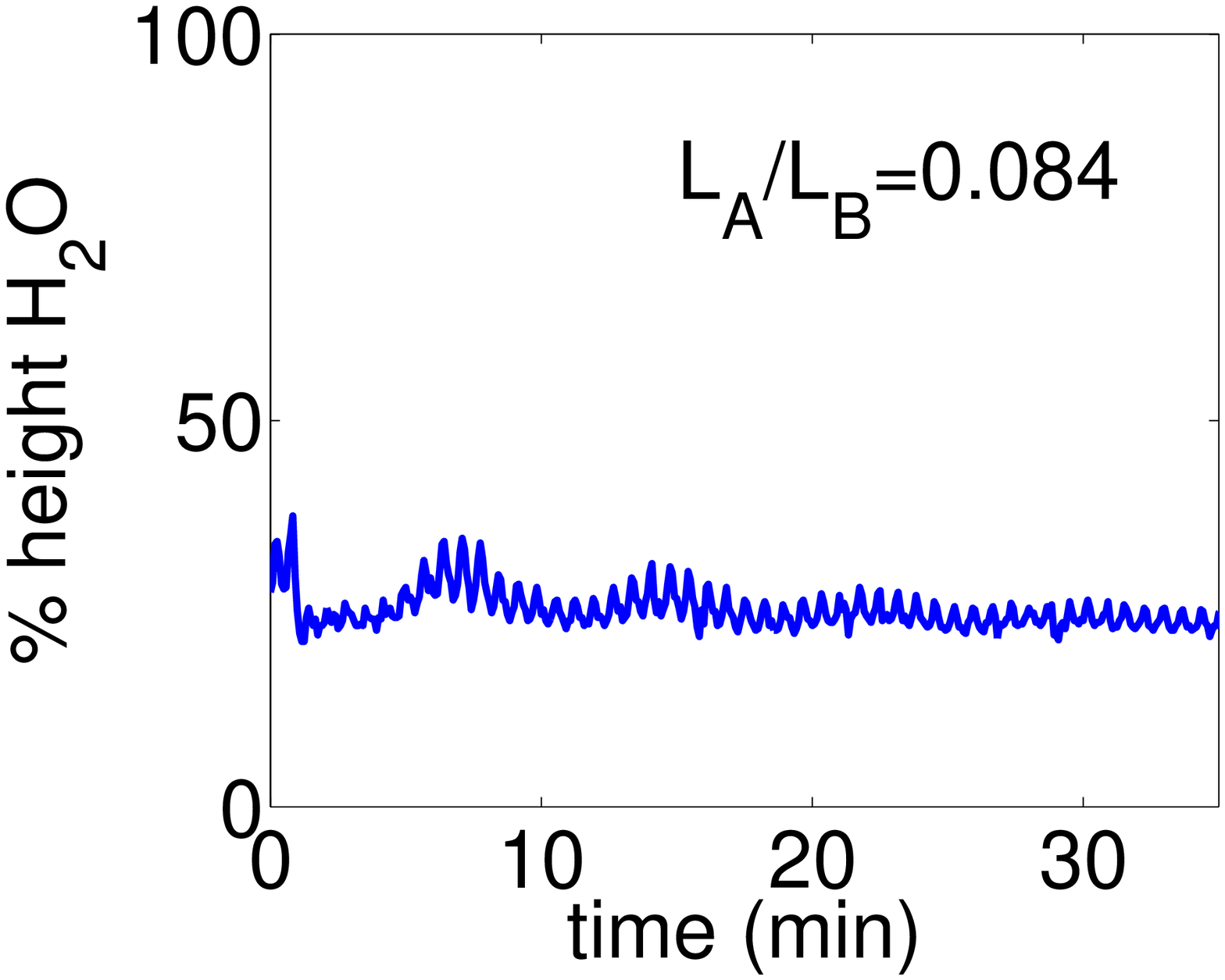}\\
\bigskip
h) \includegraphics[width=1.5in]{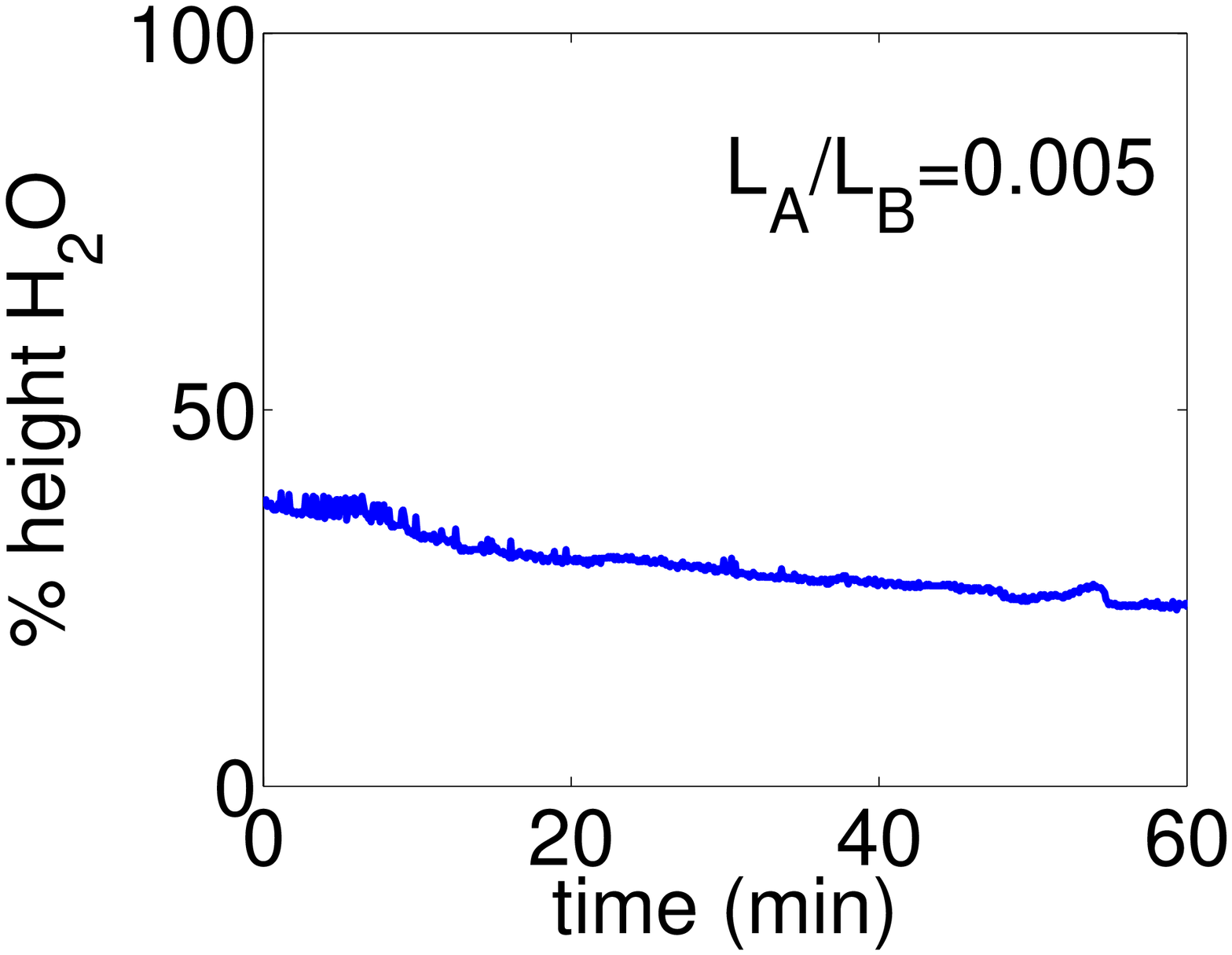}
i) \includegraphics[width=1.5in]{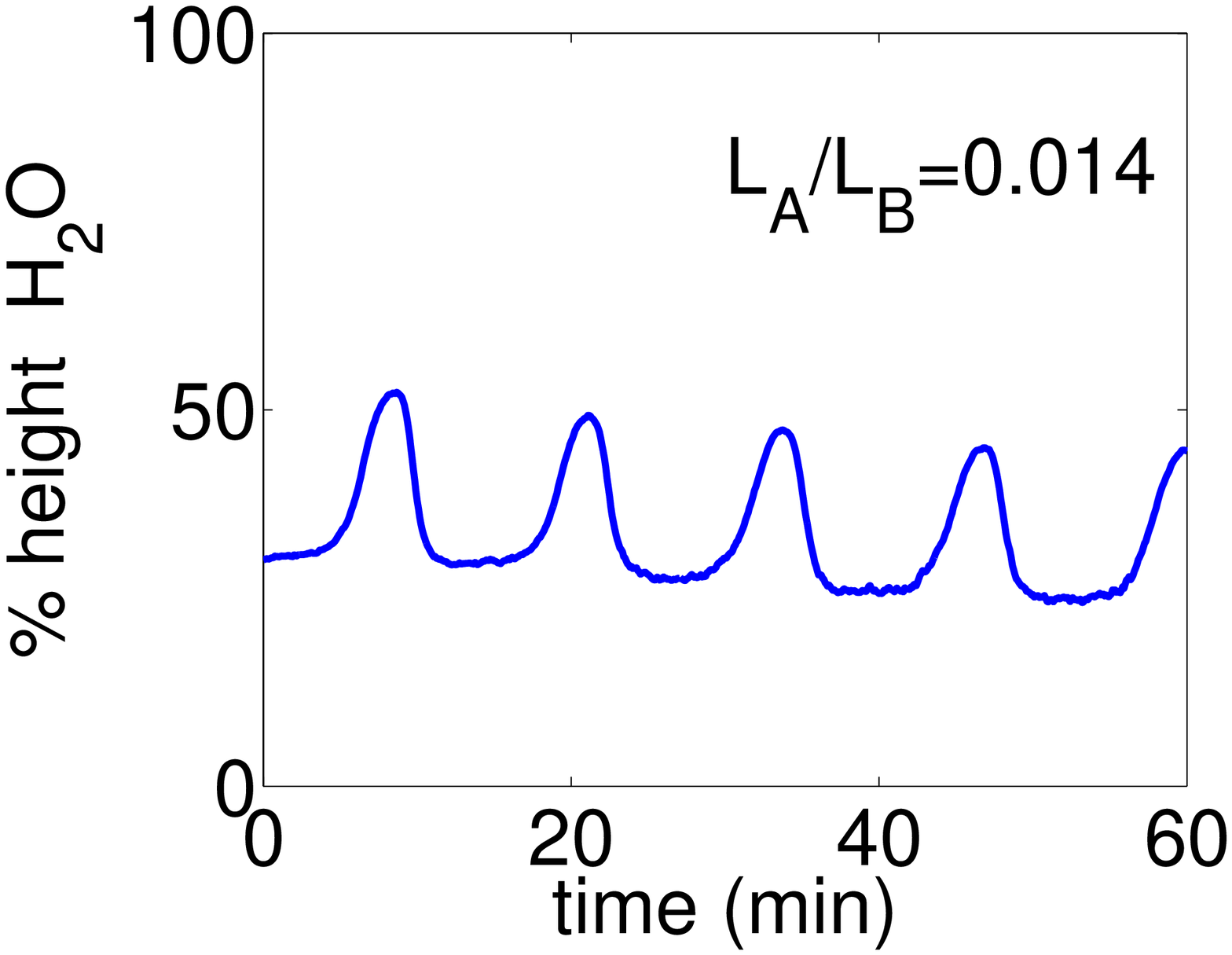}
j) \includegraphics[width=1.5in]{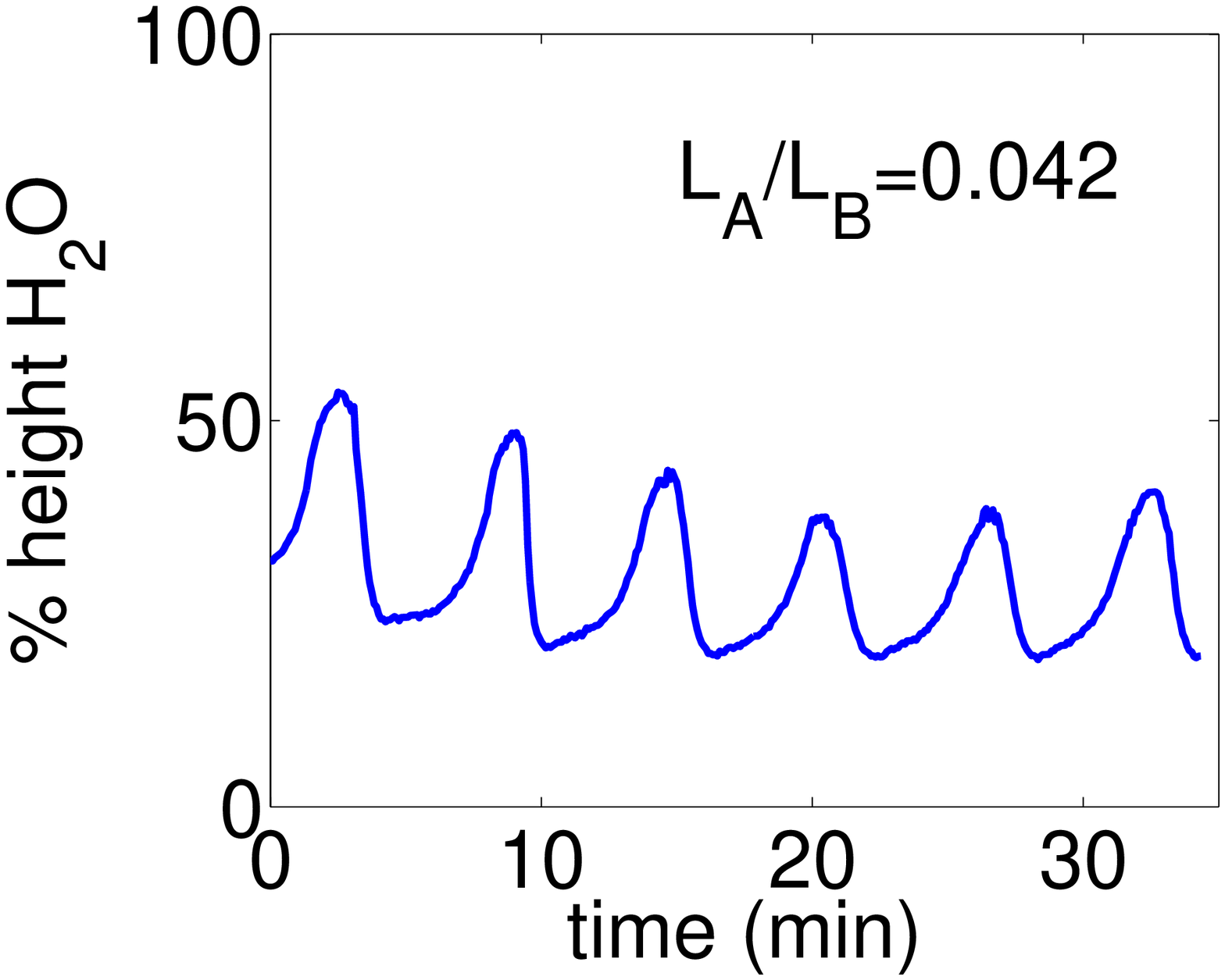}
k) \includegraphics[width=1.5in]{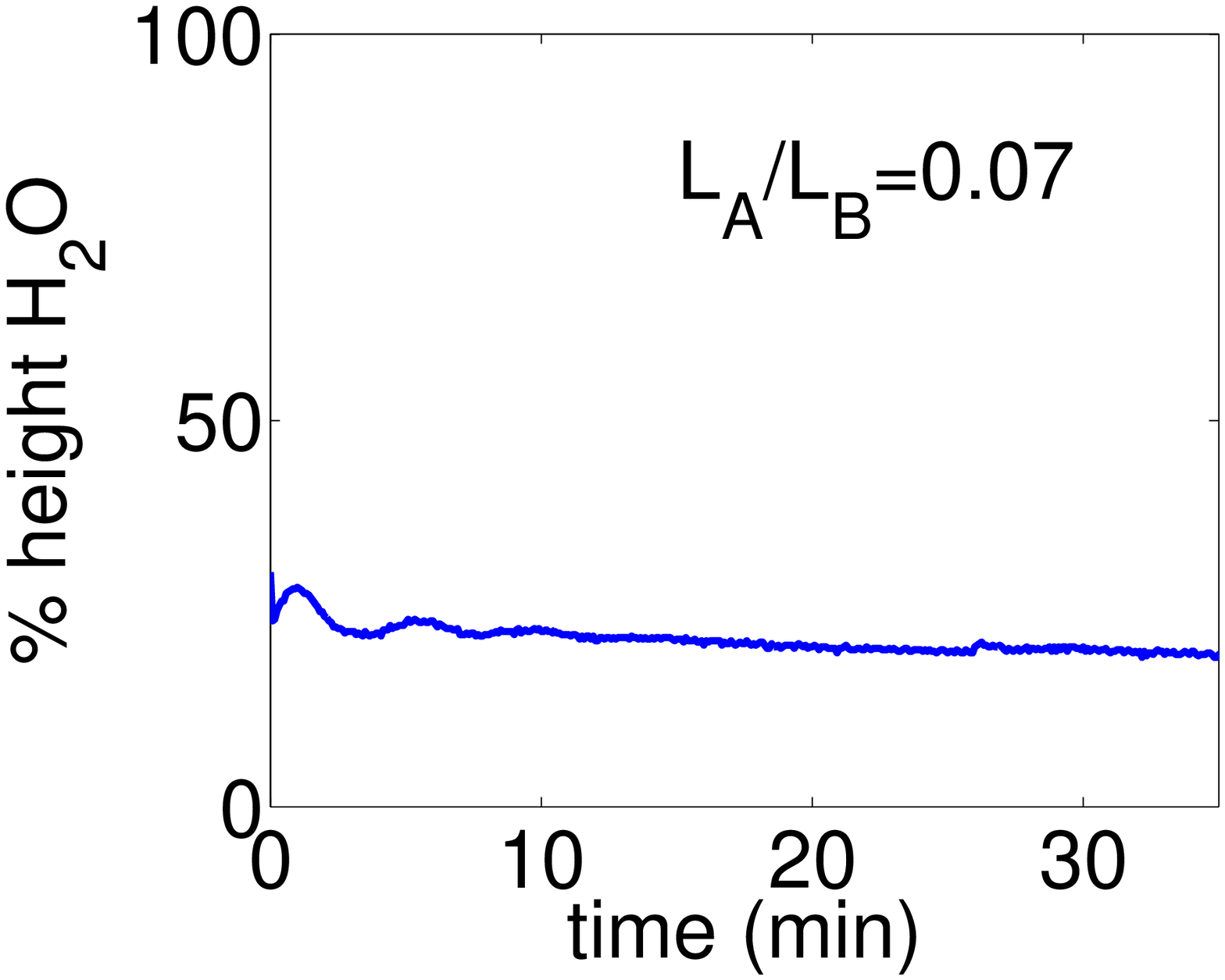}\\
\caption{a) Raw data for pressure/flow experiment where a pump is placed on the exit of vessel A or B  to control the flow in branch A. Vessels A and B are both 150 mm long and the viscous fluid is a 92\% glycerol solution by mass with a viscosity of $\mu_2=360 \mu_1$. Blue is the pressure drop across A, and red is the pressure drop across B. b) Stability function $F$ as a function of the relative flow rate in vessel A. Since the function has regions of negative slope, instability is possible in a network when $0.67<Q_A/Q_{\mathrm{in}}<0.77$ c) Stability function $F$ as a function of the length ratio. Instability corresponds to the region where this curve has positive slope, which in this case is roughly $0.004 < L_A/L_B < 0.1$. d-k) Raw time series of the location of the interface between the two fluids of the network experiments. Experiments in d-g were taken with the syringe pumps and h-k were taken with constant pressure.   }
\label{fig:criteria}
\end{figure*}

\section{Results}
This analysis provides a simple methodology to predict the stability from experimental data
since the stability function $F$ can be measured directly.
The exit network connection at vessels A and B is removed and an additional syringe pump is placed on the exit of tube A (or B) to withdraw fluid at a controlled rate. As the inlet pumps are held steady at 1 ml/min each, the outlet flow rate is varied and we measure the pressure drop across both tubes A (blue) and B (red). We then plot this experimental data as shown in Fig.~\ref{fig:criteria}a. The unusual shape of the curves is due to phase separation function at the network inlet. With no phase separation, and linear viscosity the plot would look like two straight lines of opposite slope. Rather than determining the phase separation and effective viscosity functions separately through experiment, we determine their combination in a single experiment. The flat region in the pressure-flow curves ($Q_A/Q_{\mathrm{in}}>0.77$) correspond to vessels B and C essentially containing all water~\cite{karst2013}.

From the experimental data we can directly compute the stability function $F$ and determine whether it has a region where the slope is negative, Fig.~\ref{fig:criteria}b. The locations of the maximum and minimum tell us the {\em equilibrium} flow rate, $Q_A$, which bounds the region of instability in the network; $0.67<Q_A/Q_{\mathrm{in}}<0.77$ in this case.

When the exit pump is removed and vessels A and B are connected as a network loop, the equilibrium state of the network is given by the intersection of the two pressure-flow curves in Fig.~\ref{fig:criteria}a. The equilibrium state occurs where the pressure drop in each tube is the same for a given $Q_A$. In this example where the lengths of A and B are equal, $Q_A/Q_{\mathrm{in}}\approx0.6$ and the network would be outside the instability region. However, we expect the experimental pressure drop data to scale linearly with tube length. Thus if the length of vessel B is increased, the intersection point moves to the right and into the region of instability. If we take our pressure drop data with equal length vessels, then for each value of $Q_A$, the length ratio to give that flow rate is simply $L_A/L_B = \Delta P_B/\Delta P_A$. Our data plotted in this manner is shown in Fig.~\ref{fig:criteria}c. Under this transformation,  regions with positive slope are  unstable.
Fig.~\ref{fig:criteria}c describes the stability behavior for this network for all length ratios.
At this viscosity the critical length ratios for instability are approximately $0.004<L_A/L_B <0.1$.

We  tested this analysis by building networks with different lengths and monitoring their stability. Some sample results of time series are shown in Fig.~\ref{fig:criteria} d-k. In some of the experiments, we observe that the interface between the water and glycerol solution oscillates in a regular pattern after an initial transient. While we only show snapshots here, we have recorded steady oscillations maintained for over 4 hours with a regular  $\sim5$ minute period.  In the flow controlled data the pump noise is clearly seen, however the low frequency and large amplitude oscillations of the network are unmistakeable. Under pressure control, the data are smoother, however a gradual drift is observed as the reservoirs drain and  the overall applied pressure  decreases slightly over time. The stability observations agree well with the analysis. When we are solidly in the region of instability predicted by the model, the network robustly shows spontaneous, sustained oscillations. The observed instability region is, however,  slightly narrower than that predicted by our simple model.

\begin{figure}
\begin{center}
\includegraphics[width=3in]{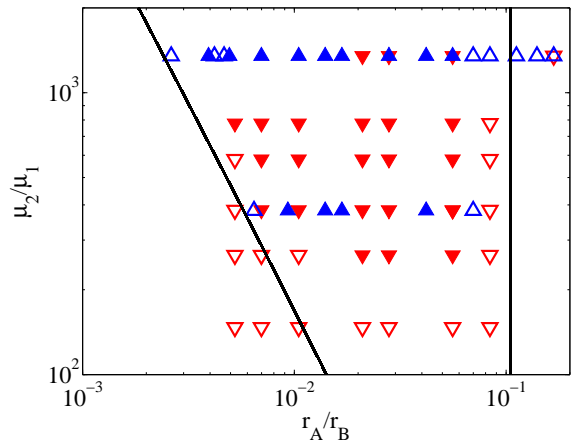}
\end{center}
\caption{Region of instability in the space of $r_A/r_B$ and $\mu_2/\mu_1$. The filled downward triangles are experimental points where we observed oscillations with flow control, the filled upward pointing triangles were oscillatory with pressure control, the open triangles are where we observed steady flow in the network (downward is flow control and upward is pressure control).  The solid line is the instability boundary predicted by Eqs. \ref{eq:left} and \ref{eq:right}}
\label{fig:ra_rb_mu}
\end{figure}

We repeat the network experiment over a range of tube lengths and a range of viscosity ratios. These data are shown in Fig.~\ref{fig:ra_rb_mu}. We find a distinct region of instability which is repeatable over a range of parameters under both pressure and flow control. Fig.~\ref{fig:ra_rb_mu} captures the results of over 60 unique experiments. We can understand this phase diagram using the iterative map. We assume that the pressure-flow curves shown in Fig.~\ref{fig:criteria} can be approximated as a linear function between $0<Q_A/Q_{\mathrm{in}}<Q^*$; where $Q^*$ is the location of the abrupt change in slope. For our data $Q^* \approx 0.8$. For $Q_A/Q_{\mathrm{in}}>Q^*$ we assume that $\Delta P_A$ is a constant and
for vessel B, we assume that it's volume fraction is zero and only water is in that vessel.


Using this empirical linear model and taking the limit as the viscosity ratio becomes large,
 yields a simple approximation.
With these assumptions the critical values of the equilibrium flow rate
are $Q^*$ and $Q^*/(2-Q^*)$ and the iterative map gives the critical value of $r_A/r_B$ on the left boundary as,
\begin{equation}
\label{eq:left}
\frac{r_A}{r_B} = \frac{1-Q^*}{ \mu\left( {\frac{\mu_2}{\mu_1},\Phi_\textrm{in}},d_A\right)},
\end{equation}
and on the right boundary as,
\begin{equation}
\label{eq:right}
\frac{r_A}{r_B} = \left(1-Q^*\right) \frac{\mu\left(\frac{\mu_2}{\mu_1},\Phi_{\textrm{in}},d_B\right)}
                                          {\mu\left(\frac{\mu_2}{\mu_1},\Phi_{\textrm{in}},d_A\right)}.
\end{equation}
These boundaries are shown, assuming $Q^*=0.8$ for all viscosities, as the solid curves in Fig.~\ref{fig:ra_rb_mu} which capture the general trend of the data.
We obtained an approximate effective viscosity function for
our stratified flow from measuring the pressure drop in single tubes.
Note  that we  measure the effective viscosity in vessel B to be
about half the effective viscosity as vessel A for the same volume fraction due to vessel B's smaller diameter.
 Our observations show that the instability disappears at viscosity contrasts below $\mu_2/\mu_1 < 200$.  Previous experiments show that as the viscosity contrast is reduced  the critical point $Q^*\rightarrow 1$ and eventually disappears~\cite{karst2013}. Thus, the assumption that $Q^*$ is constant is not true, however the basic trends are captured by this simple model. The stability observations indicate the critical boundary  on the right is insensitive to viscosity contrast
while the left boundary has a slope inversely proportional to the viscosity.

\section{Conclusions}
We have demonstrated  a simple piping network system that shows spontaneous and sustained oscillations over a wide range of parameters. The geometry of the network is perhaps one of the simplest under which  such oscillations could exist. The network only has one diverging node at the inlet and only one internal degree of freedom - the distribution of flow between the two halves of the loop.
The oscillations have been observed under both pressure and flow control and robustly exist over a wide range of experiments. The mechanism of the oscillations is easily understood with an iterative map where a single tube in the system introduces a delay for changes which propagate through the system.  Since the fluid flow in the system adjusts itself instantly, and everywhere within the network loop at once, there is a delay in the feedback mechanism to adjust the flow.

In this paper, our results are applied to a model  system.
However,  the  analysis provides a simple way to probe whether such
instabilities could be observed in other physical systems.
 If one has access to experimentally obtained phase separation and effective viscosity data, one can easily use Eq. 4 to determine whether this type of instability could be observed.
 We note that other distinct types of oscillations may exist, such as high frequency ones where the assumptions of the iterative map breakdown, which are not captured by the simple criteria ~\cite{karst2014}.

In microvascular blood flow,  network oscillations  have been observed experimentally in complex geometries and predicted theoretically in simple ones. These predictions often lie in ranges of parameter space which are not experimentally testable. Our results demonstrate that  spontaneous oscillations  can emerge in simple network geometries and we provide a laboratory system where theory and experiment are in excellent agreement.

\section{Acknowledgements}
This work was supported by the National Science
Foundation under Contract No. DMS-1211640. We thank Erika Weiler who assisted with some early experiments.

\bibliography{refs_network}

\end{document}